\documentclass[10pt,a4paper]{article}
\usepackage{amsfonts,amsmath,amssymb,cite}
\usepackage[colorlinks=true,citecolor=black,linkcolor=black,%
urlcolor=blue]{hyperref}
\begin{document}
\title{An analogue of the P{\"o}schl--Teller anharmonic oscillator \\
on an $N$-dimensional sphere}
\author{Rados{\l}aw Szmytkowski\footnote{Email:
radoslaw.szmytkowski@pg.edu.pl} \\*[3ex]
Faculty of Applied Physics and Mathematics,
Gda{\'n}sk University of Technology, \\
ul.\ Gabriela Narutowicza 11/12, 80--233 Gda{\'n}sk, Poland}
\date{\today}
\maketitle
\begin{abstract}
A Schr{\"o}dinger particle on an $N$-dimensional ($N\geqslant2$)
hypersphere of radius $R$ is considered. The particle is subjected to
the action of a force characterized by the potential
$V(\theta)=2m\omega_{1}^{2}R^{2}\tan^{2}(\theta/2)
+2m\omega_{2}^{2}R^{2}\cot^{2}(\theta/2)$, where
$0\leqslant\theta\leqslant\pi$ is the hyperlatitude angular
coordinate. In the general case when $\omega_{1}\neq\omega_{2}$, this
is a model of a hyperspherical analogue of the P{\"o}schl--Teller
anharmonic oscillator. Energy eigenvalues and normalized
eigenfunctions for this system are found in closed analytical forms.
For $N=2$, our results reproduce those obtained by Kazaryan \emph{et
al.\/} [\href{https://doi.org/10.1016/j.physe.2013.03.014}{Physica E
52 (2013) 122}]. For $N\geqslant2$ arbitrary and for $\omega_{2}=0$,
the results of Mardoyan and Petrosyan
[\href{https://doi.org/10.3103/S1068337213020035}%
{J.\/~Contemp.\/~Phys.\ 48 (2013) 70}] for their model of an isotropic
hyperspherical harmonic oscillator are recovered. The Euclidean limit
for the anharmonic oscillator in question is also discussed. \\*[3ex]
\noindent 
\textbf{Key words:} P{\"o}schl--Teller oscillator; spherical geometry;
Schr{\"o}dinger equation; Jacobi polynomials; Gegenbauer polynomials;
stereographic projection
\end{abstract}
%
%
\section{Introduction}
\label{I}
\setcounter{equation}{0}
The search for closed-form analytical solutions to quantum-mechanical
wave equations in spherical geometry dates back at least to the early
1940s. It was then that Schr{\"o}dinger \cite{Schr40} and Stevenson
\cite{Stev41} considered an analogue of the Schr{\"o}dinger--Coulomb
energy eigenvalue problem on the hypersphere $\mathbb{S}_{R}^{3}$,
with the potential $V(\theta)=V_{0}\cot\theta$, where
$0\leqslant\theta\leqslant\pi$ is the hyperlatitude angle on
$\mathbb{S}_{R}^{3}$. At the end of the 1970s, Higgs \cite{Higg79} and
Leemon \cite{Leem79} solved the spherical Schr{\"o}dinger equation for
the hemispherical analogue of the isotropic harmonic oscillator with
the latitudinal potential $V(\theta)=V_{0}\tan^{2}\!\theta$
($V_{0}\geqslant0$, $0\leqslant\theta\leqslant\pi/2$). The two systems
mentioned, like their counterparts in the Euclidean space, possess
hidden dynamical symmetries. For this reason, they appeared to be
attractive objects of research and in the following years their
various properties were comprehensively studied in a number of works.
The most important results obtained in the course of research
conducted up to the mid- and late-2000s were reviewed in the
monographs by Shchepetilov \cite{Shch06} and by Redkov and Ovsiyuk
\cite{Redk12}, respectively.

In 2013, a paper by Mardoyan and Petrosyan \cite{Mard13} was
published, in which they found analytical solutions to the energy
eigenproblem on the hypersphere $\mathbb{S}_{R}^{N}$ for a
Schr{\"o}dinger particle in the latitudinal potential
$V(\theta)=V_{0}\tan^{2}(\theta/2)$, ($V_{0}\geqslant0$,
$0\leqslant\theta\leqslant\pi$). These authors also showed that in the
Euclidean limit the system considered by them, similarly to the one
discussed earlier by Higgs \cite{Higg79} and Leemon \cite{Leem79},
passes into an $N$-dimensional isotropic harmonic oscillator.

Also in 2013, Kazaryan \emph{et al.\/} \cite{Kaza13} considered a
Schr{\"o}dinger particle moving on a two-dimensional sphere
$\mathbb{S}_{R}^{2}$, in the potential $V(\theta)
=V_{1}\tan^{2}(\theta/2)+V_{2}\cot^{2}(\theta/2)$, ($V_{1}\geqslant0$,
$V_{2}\geqslant0$, $0\leqslant\theta\leqslant\pi$), and found
analytical solutions to the corresponding energy eigenproblem. In the
case of $V_{2}=0$ and $V_{1}>0$ (and, due to the invariance of the
potential in question with respect to the combined transformations
$\theta\leftrightarrow\pi-\theta$ and $V_{1}\leftrightarrow V_{2}$,
also in the case of $V_{1}=0$ and $V_{2}>0$), the system considered
in Ref.\ \cite{Kaza13} evidently reduces to the two-dimensional
example of a spherical isotropic harmonic oscillator of Mardoyan and
Petrosyan \cite{Mard13}. On the other hand, in the case when both
$V_{1}$ and $V_{2}$ are different from zero, the system from Ref.\
\cite{Kaza13} may be seen as a kind of a (two-dimensional) spherical
analogue of the anharmonic P{\"o}schl--Teller oscillator
\cite{Posc33}.

The purpose of this work is to generalize the results from Ref.\
\cite{Kaza13} to the case when the domain is an $N$-dimensional
hypersphere $\mathbb{S}_{R}^{N}$, ($N\geqslant2$). The paper is
structured as follows. In Sec.\ \ref{II}, we recall the basic relevant
facts from the methodology of treating the Schr{\"o}dinger equation in
the spherical geometry. In particular, we give a quasi-radial
Schr{\"o}dinger equation for potentials which depend on the
hyperlatitudinal angle only. In Sec.\ \ref{III}, the Sturm--Liouville
problem involving such an equation with a potential describing a
spherical analogue of the P{\"o}schl--Teller anharmonic oscillator is
solved, yielding energy eigenvalues and associated normalized
quasi-radial eigenfunctions in closed analytical forms. Some special
cases are then considered in more detail in Sec.\ \ref{IV}. In Sec.\
\ref{V}, we study the Euclidean limit for the oscillator in question
by stereographically projecting the system onto a suitably chosen
tangent space and then by going with the radius of the sphere to
infinity. The main results of the work are summarized in Sec.\
\ref{VI}.
%
%
\section{The Schr{\"o}dinger equation in a spherical geometry}
\label{II}
\setcounter{equation}{0}
Let $\mathbb{S}_{R}^{N}\subset\mathbb{R}^{N+1}$ be an $N$-dimensional
hypersphere with its center at the point
$\mathbf{0}=(0,\ldots,0)\in\mathbb{R}^{N+1}$ and of radius $R$. Its
equation is
\begin{equation}
\sum_{n=1}^{N+1}x_{n}^{2}=R^{2},
\label{2.1}
\end{equation}
where $\{x_{n}\}_{n=1}^{N+1}$ are coordinates in some Cartesian system
in $\mathbb{R}^{N+1}$ with its origin at $\mathbf{0}$. The position of
a point on $\mathbb{S}_{R}^{N}$ may be uniquely determined by giving
its hyperspherical angular coordinates $\{\theta_{n}\}_{n=1}^{N}$
related to the Cartesian coordinates $\{x_{n}\}_{n=1}^{N+1}$ as
follows:
\begin{subequations}
\begin{equation}
\left\{
\begin{array}{rcl}
x_{1} \!\!\!\! &=& \!\!\!\! R\sin\theta_{N}\sin\theta_{N-1}
\ldots\sin\theta_{3}\sin\theta_{2}\cos\theta_{1},
\\
x_{2} \!\!\!\! &=& \!\!\!\! R\sin\theta_{N}\sin\theta_{N-1}
\ldots\sin\theta_{3}\sin\theta_{2}\sin\theta_{1},
\\
x_{3} \!\!\!\! &=& \!\!\!\! R\sin\theta_{N}\sin\theta_{N-1}
\ldots\sin\theta_{3}\cos\theta_{2}, \\
x_{4} \!\!\!\! &=& \!\!\!\! R\sin\theta_{N}\sin\theta_{N-1}
\ldots\cos\theta_{3}, \\
&\vdots& \\
x_{N} \!\!\!\! &=& \!\!\!\! R\sin\theta_{N}\cos\theta_{N-1}, \\
x_{N+1} \!\!\!\! &=&  \!\!\!\! R\cos\theta_{N},
\end{array}
\right.
\label{2.2a}
\end{equation}
with
\begin{equation}
0\leqslant\theta_{1}<2\pi
\label{2.2b}
\end{equation}
and
\begin{equation}
0\leqslant\theta_{n}\leqslant\pi
\qquad (2\leqslant n\leqslant N).
\label{2.2c}
\end{equation}
\label{2.2}%
\end{subequations}
The coordinate $\theta_{N}$ is termed the hyperlatitude or the
quasi-radial variable. The points for which $\theta_{N}=0$ or
$\theta=\pi$ are sometimes called the \emph{north\/} or the
\emph{south\/} poles of the hypersphere, respectively. In what
follows, the set of the coordinates $\{\theta_{n}\}_{n=1}^{N}$ as an
entity will be denoted as $\Omega_{N}$; similarly, hereafter the
symbol $\Omega_{N-1}$ will stand for the set
$\{\theta_{n}\}_{n=1}^{N-1}$.

The Laplace--Beltrami operator on $\mathbb{S}_{R}^{N}$ is related to
its counterpart on the unit sphere $\mathbb{S}^{N}$ simply through
\begin{equation}
\Delta_{\mathbb{S}_{R}^{N}}=\frac{1}{R^{2}}\Delta_{\mathbb{S}^{N}}.
\label{2.3}
\end{equation}
The latter operator may be conveniently defined by the chain of
recurrence relations
\begin{subequations}
\begin{align}
\Delta_{\mathbb{S}^{n}}
&= \frac{1}{\sin^{n-1}\!\theta_{n}}\frac{\partial}{\partial\theta_{n}}
\sin^{n-1}\!\theta_{n}\frac{\partial}{\partial\theta_{n}}
+\frac{1}{\sin^{2}\!\theta_{n}}\Delta_{\mathbb{S}^{n-1}}
\nonumber \\
&= \frac{\partial^{2}}{\partial\theta_{n}^{2}}
+(n-1)\cot\theta_{n}\frac{\partial}{\partial\theta_{n}}
+\frac{1}{\sin^{2}\!\theta_{n}}\Delta_{\mathbb{S}^{n-1}}
\qquad (2\leqslant n\leqslant N),
\label{2.4a}
\end{align}
with
\begin{equation}
\Delta_{\mathbb{S}^{1}}
=\frac{\partial^{2}}{\partial\theta_{1}^{2}}.
\label{2.4b}
\end{equation}
\label{2.4}%
\end{subequations}

The time-independent Schr{\"o}dinger partial differential equation for
a particle moving on the sphere $\mathbb{S}_{R}^{N}$, ($N\geqslant2$),
under the action of a force derivable from the potential
$V(\Omega_{N})$, is
\begin{equation}
\left[-\frac{\hbar^{2}}{2mR^{2}}\Delta_{\mathbb{S}^{N}}
+V(\Omega_{N})\right]\Psi(\Omega_{N})=E\Psi(\Omega_{N}).
\label{2.5}
\end{equation}
In the special case when the potential $V$ is a function of the
hyperlatitude only, i.e., when it holds that
\begin{equation}
V(\Omega_{N})=V(\theta_{N}),
\label{2.6}
\end{equation}
Eq.\ (\ref{2.5}) is separable and admits particular solutions that are
of the form
\begin{equation}
\Psi_{L\eta}(\Omega_{N})
=F_{L}(\theta_{N})Y_{L\eta}^{(N-1)}(\Omega_{N-1}).
\label{2.7}
\end{equation}
Here $Y_{L\eta}^{(N-1)}(\Omega_{N-1})$ are the hyperspherical
harmonics, i.e., the eigenfunctions of the Laplace--Beltrami operator
on $\mathbb{S}^{N-1}$,
\begin{equation}
\Delta_{\mathbb{S}^{N-1}}Y_{L\eta}^{(N-1)}(\Omega_{N-1})
=-L(L+N-2)Y_{L\eta}^{(N-1)}(\Omega_{N-1}),
\label{2.8}
\end{equation}
which are normalizable in the sense of
\begin{equation}
\oint_{\mathbb{S}^{N-1}}\mathrm{d}^{N-1}\Omega_{N-1}\:
\big|Y_{L\eta}^{(N-1)}(\Omega_{N-1})\big|^{2}=1.
\label{2.9}
\end{equation}
We take $L\in\mathbb{Z}$ for $N=2$ and $L\in\mathbb{N}_{0}$ for
$N\geqslant3$. If $N=2$, then $\eta$ is redundant and is to be
suppressed, while if $N\geqslant3$, then $\eta$ stands collectively
for all quantum numbers necessary to distinguish between the
hyperspherical harmonics that belong to the (degenerate for $L\neq0$)
eigenvalue $-L(L+N-2)$ of $\Delta_{\mathbb{S}^{N-1}}$. The function
$F_{L}(\theta_{N})$ obeys the quasi-radial Schr{\"o}dinger equation
\begin{equation}
\left\{-\frac{\hbar^{2}}{2mR^{2}}
\left[\frac{\mathrm{d}^{2}}{\mathrm{d}\theta_{N}^{2}}
+(N-1)\cot\theta_{N}\frac{\mathrm{d}}{\mathrm{d}\theta_{N}}
-\frac{L(L+N-2)}{\sin^{2}\!\theta_{N}}\right]
+V(\theta_{N})-E\right\}F_{L}(\theta_{N})=0.
\label{2.10}
\end{equation}

On the following pages, we shall derive and analyze solutions to Eq.\
(\ref{2.10}) with arbitrary $N\geqslant2$ and with the potential
\begin{subequations}
\begin{equation}
V(\theta_{N})=2m\omega_{1}^{2}R^{2}\tan^{2}\!\frac{\theta_{N}}{2}
+2m\omega_{2}^{2}R^{2}\cot^{2}\!\frac{\theta_{N}}{2}
\qquad (0\leqslant\theta_{N}\leqslant\pi),
\label{2.11a}
\end{equation}
where, for definiteness, we shall be assuming that
$\omega_{1}\geqslant0$, $\omega_{2}\geqslant0$. The system
characterized by the above potential is essentially a hyperspherical
analogue of the P{\"o}schl--Teller \cite{Posc33} anharmonic
oscillator. This is justified by the fact that after a bit of
trigonometry Eq.\ (\ref{2.11a}) may be rewritten in the form
\begin{equation}
V(\theta_{N})=\frac{2m\omega_{1}^{2}R^{2}}{\cos^{2}(\theta_{N}/2)}
+\frac{2m\omega_{2}^{2}R^{2}}{\sin^{2}(\theta_{N}/2)}
-2m\big(\omega_{1}^{2}+\omega_{2}^{2}\big)R^{2}
\qquad (0\leqslant\theta_{N}\leqslant\pi).
\label{2.11b}
\end{equation}
\label{2.11}%
\end{subequations}
Once Eq.\ (\ref{2.11b}) is plugged into Eq.\ (\ref{2.10}) and the
formal replacements $N\mapsto1$, $L\mapsto0$ and $E\mapsto
E-2m(\omega_{1}^{2}+\omega_{2}^{2})R^{2}$ are made therein, the
resulting equation is found to coincide, up to notational differences,
with the one-dimensional Schr{\"o}dinger equation following from Eqs.\
(2a) and (3) of Ref.\ \cite{Posc33}.

Before moving on, it should be emphasized that the \emph{spherical\/}
analogue of the P{\"o}schl--Teller oscillator considered in this work
differs from the system with the potential
\begin{align}
V(\theta_{N}) &= \frac{1}{2}m\omega_{1}^{2}R^{2}\tan^{2}\theta_{N}
+\frac{1}{2}m\omega_{2}^{2}R^{2}\cot^{2}\theta_{N}
\nonumber \\
&= \frac{m\omega_{1}^{2}R^{2}}{2\cos^{2}\theta_{N}}
+\frac{m\omega_{2}^{2}R^{2}}{2\sin^{2}\theta_{N}}
-\frac{1}{2}m\big(\omega_{1}^{2}+\omega_{2}^{2}\big)R^{2}
\qquad (0\leqslant\theta_{N}\leqslant\pi/2),
\label{2.12}
\end{align}
which also admits analytical solutions to the quasi-radial
Schr{\"o}dinger equation (\ref{2.10}), albeit on a different domain,
and which in turn can be viewed as a \emph{hemispherical\/} analogue
of the P{\"o}schl--Teller oscillator.

For the sake of notational clarity, from now on we shall be dropping
the index $N$ at $\theta_{N}$.
%
%
\section{Energy levels and normalized eigenfunctions for the potential
(\ref{2.11a}) on an $N$-dimensional sphere}
\label{III}
\setcounter{equation}{0}
The quasi-radial equation (\ref{2.10}) with the potential
(\ref{2.11a}) is
\begin{align}
\Bigg[\frac{\mathrm{d}^{2}}{\mathrm{d}\theta^{2}}
+(N-1)\cot\theta\frac{\mathrm{d}}{\mathrm{d}\theta}
-\frac{L(L+N-2)}{\sin^{2}\!\theta}
-\frac{4m^{2}\omega_{1}^{2}R^{4}}{\hbar^{2}}\tan^{2}\!\frac{\theta}{2}
-\frac{4m^{2}\omega_{2}^{2}R^{4}}{\hbar^{2}}\cot^{2}\!\frac{\theta}{2}
+\frac{2mR^{2}}{\hbar^{2}}E\Bigg]F_{L}(\theta)=0. &
\nonumber \\
&
\label{3.1}
\end{align}
We wish to solve it under the constraint that $F_{L}(\theta)$ remains
bounded as $\theta\to0+0$ and as $\theta\to\pi-0$, considering the
energy parameter $E$ as an eigenvalue.

In the first step, we switch from the variable $\theta$ to the new
one:
\begin{equation}
\rho=\cos\theta
\qquad (-1\leqslant\rho\leqslant1)
\label{3.2}
\end{equation}
and make the substitution
\begin{equation}
F_{L}(\theta)\equiv F_{L}(\rho)
=\frac{f_{L}(\rho)}{(1-\rho^{2})^{N/4-1/2}}.
\label{3.3}
\end{equation}
This casts Eq.\ (\ref{3.1}) into the following one:
\begin{equation}
\left[\frac{\mathrm{d}}{\mathrm{d}\rho}(1-\rho^{2})
\frac{\mathrm{d}}{\mathrm{d}\rho}+\lambda(\lambda+1)
-\frac{\mu_{L2}^{2}}{2(1-\rho)}
-\frac{\mu_{L1}^{2}}{2(1+\rho)}\right]f_{L}(\rho)=0,
\label{3.4}
\end{equation}
with
\begin{equation}
\lambda=-\frac{1}{2}+\frac{1}{2}\sqrt{(N-1)^{2}
+\frac{16m^{2}\big(\omega_{1}^{2}+\omega_{2}^{2}\big)R^{4}}{\hbar^{2}}
+\frac{8mR^{2}}{\hbar^{2}}E}
\label{3.5}
\end{equation}
and
\begin{equation}
\mu_{Lk}=\sqrt{\left(L+\frac{N}{2}-1\right)^{2}
+\left(\frac{4m\omega_{k}R^{2}}{\hbar}\right)^{2}}
\qquad (k=1,2).
\label{3.6}
\end{equation}
Equation (\ref{3.4}) is the generalized associated Legendre equation
\cite{Kuip57a,Kuip57b,Virc01}. Its general solution is
\begin{equation}
f_{L}(\rho)=A_{L}P_{\lambda}^{-\mu_{L2},\mu_{L1}}(\rho)
+B_{L}P_{\lambda}^{\mu_{L2},\mu_{L1}}(\rho),
\label{3.7}
\end{equation}
where
\begin{equation}
P_{\lambda}^{\mu,\nu}(\rho)=\frac{1}{\Gamma(1-\mu)}
\frac{(1+\rho)^{\nu/2}}{(1-\rho)^{\mu/2}}\,
{}_{2}F_{1}\left(-\lambda-\frac{\mu-\nu}{2},
\lambda+1-\frac{\mu-\nu}{2};1-\mu;\frac{1-\rho}{2}\right)
\label{3.8}
\end{equation}
is the generalized associated Legendre function of the first kind,
while $A_{L}$ and $B_{L}$ are arbitrary constants. Combining Eqs.\
(\ref{3.3}), (\ref{3.7}) and (\ref{3.8}) yields the quasi-radial
function $F_{L}(\rho)$ in terms of the hypergeometric function
${}_{2}F_{1}$:
\begin{align}
F_{L}(\rho) &= \;A_{L}^{\prime}
(1-\rho)^{\mu_{L2}/2-N/4+1/2}(1+\rho)^{\mu_{L1}/2-N/4+1/2}
\nonumber \\
&\qquad \times{}_{2}F_{1}\left(-\lambda+\frac{\mu_{L1}+\mu_{L2}}{2},
\lambda+1
+\frac{\mu_{L1}+\mu_{L2}}{2};1+\mu_{L2};\frac{1-\rho}{2}\right)
\nonumber \\
& \quad +B_{L}^{\prime}(1-\rho)^{-\mu_{L2}/2-N/4+1/2}
(1+\rho)^{\mu_{L1}/2-N/4+1/2}
\nonumber \\
& \qquad
\times{}_{2}F_{1}\left(-\lambda+\frac{\mu_{L1}-\mu_{L2}}{2},
\lambda+1
+\frac{\mu_{L1}-\mu_{L2}}{2};1-\mu_{L2};\frac{1-\rho}{2}\right),
\label{3.9}
\end{align}
with $A_{L}^{\prime}$ and $B_{L}^{\prime}$ being arbitrary constants.
The presence of the factor $(1-\rho)^{-\mu_{L2}/2-N/4+1/2}$ implies
that the second term on the right-hand side of Eq.\ (\ref{3.9})
diverges as $\rho\to1-0$. To get rid of this term, we set
$B_{L}^{\prime}=0$, thus obtaining
\begin{align}
F_{L}(\rho) &= \;A_{L}^{\prime}
(1-\rho)^{\mu_{L2}/2-N/4+1/2}(1+\rho)^{\mu_{L1}/2-N/4+1/2}
\nonumber \\
&\qquad \times{}_{2}F_{1}\left(-\lambda+\frac{\mu_{L1}+\mu_{L2}}{2},
\lambda+1
+\frac{\mu_{L1}+\mu_{L2}}{2};1+\mu_{L2};\frac{1-\rho}{2}\right).
\label{3.10}
\end{align}
The hypergeometric function in Eq.\ (\ref{3.10}) diverges as
$\rho\to-1+0$ unless we stipulate that
\begin{equation}
-\lambda+\frac{\mu_{L1}+\mu_{L2}}{2}=-n_{\theta}
\qquad (n_{\theta}\in\mathbb{N}_{0}).
\label{3.11}
\end{equation}
Merging this condition with the definition (\ref{3.5}) of $\lambda$
gives the quantized energy levels for the system under the
consideration:
\begin{subequations}
\begin{align}
E_{n_{\theta}L} &= \frac{\hbar^{2}}{2mR^{2}}
\left[n_{\theta}+\frac{N}{2}
+\frac{1}{2}\sqrt{\left(L+\frac{N}{2}-1\right)^{2}
+\left(\frac{4m\omega_{1}R^{2}}{\hbar}\right)^{2}}
+\frac{1}{2}\sqrt{\left(L+\frac{N}{2}-1\right)^{2}
+\left(\frac{4m\omega_{2}R^{2}}{\hbar}\right)^{2}}\,\right]
\nonumber \\
& \qquad \times\left[n_{\theta}-\frac{N}{2}+1
+\frac{1}{2}\sqrt{\left(L+\frac{N}{2}-1\right)^{2}
+\left(\frac{4m\omega_{1}R^{2}}{\hbar}\right)^{2}}
+\frac{1}{2}\sqrt{\left(L+\frac{N}{2}-1\right)^{2}
+\left(\frac{4m\omega_{2}R^{2}}{\hbar}\right)^{2}}\,\right]
\nonumber \\
& \quad -2m\big(\omega_{1}^{2}+\omega_{2}^{2}\big)R^{2},
\label{3.12a}
\end{align}
or equivalently
\begin{align}
E_{n_{\theta}L} &= \frac{\hbar^{2}}{2mR^{2}}
\Bigg\{\left(n_{\theta}+\frac{N}{2}\right)
\left(n_{\theta}-\frac{N}{2}+1\right)
+\frac{1}{2}\left(L+\frac{N}{2}-1\right)^{2}
\nonumber \\
& \quad +\left(n_{\theta}+\frac{1}{2}\right)
\left[\sqrt{\left(L+\frac{N}{2}-1\right)^{2}
+\left(\frac{4m\omega_{1}R^{2}}{\hbar}\right)^{2}}
+\sqrt{\left(L+\frac{N}{2}-1\right)^{2}
+\left(\frac{4m\omega_{2}R^{2}}{\hbar}\right)^{2}}\,\right]
\nonumber \\
& \quad +\frac{1}{2}\sqrt{\left[\left(L+\frac{N}{2}-1\right)^{2}
+\left(\frac{4m\omega_{1}R^{2}}{\hbar}\right)^{2}\right]
\left[\left(L+\frac{N}{2}-1\right)^{2}
+\left(\frac{4m\omega_{2}R^{2}}{\hbar}\right)^{2}\right]}\,\Bigg\}.
\label{3.12b}
\end{align}
\label{3.12}%
\end{subequations}
For $N=2$, either of Eqs.\ (\ref{3.12}) may be shown to reduce, up to
notational differences, to Eq.\ (25) in Ref.\ \cite{Kaza13}.

To obtain the most suitable form of the quasi-radial eigenfunctions
associated with the eigenenergies (\ref{3.12}), we observe that
insertion of Eq.\ (\ref{3.11}) into Eq.\ (\ref{3.10}) gives
\begin{align}
F_{n_{\theta}L}(\rho) &= A_{L}^{\prime}
(1-\rho)^{\mu_{L2}/2-N/4+1/2}(1+\rho)^{\mu_{L1}/2-N/4+1/2}
{}_{2}F_{1}\left(-n_{\theta},n_{\theta}+\mu_{L1}+\mu_{L2}+1;
\mu_{L2}+1;\frac{1-\rho}{2}\right).
\nonumber \\
&
\label{3.13}
\end{align}
The function ${}_{2}F_{1}$ appearing in the above equation is closely
related to the Jacobi polynomial \cite[p.\/~212]{Magn66}:
\begin{equation}
P_{n_{\theta}}^{(\mu_{L2},\mu_{L1})}(\rho)
=\frac{\Gamma(n_{\theta}+\mu_{L2}+1)}{n_{\theta}!\Gamma(\mu_{L2}+1)}\,
{}_{2}F_{1}\left(-n_{\theta},n_{\theta}+\mu_{L1}+\mu_{L2}+1;
\mu_{L2}+1;\frac{1-\rho}{2}\right)
\label{3.14}
\end{equation}
[not to be confused with the generalized associated Legendre function
of the first kind defined in Eq.\ (\ref{3.8})]. Hence, we may write
\begin{equation}
F_{n_{\theta}L}(\rho) =A_{L}^{\prime\prime}
(1-\rho)^{\mu_{L2}/2-N/4+1/2}(1+\rho)^{\mu_{L1}/2-N/4+1/2}
P_{n_{\theta}}^{(\mu_{L2},\mu_{L1})}(\rho),
\label{3.15}
\end{equation}
where $A_{L}^{\prime\prime}$ is a non-zero constant. We choose the
value of the latter as to have the complete wave function (\ref{2.7})
normalized to the unity in the sense of
\begin{equation}
R^{N}\oint_{\mathbb{S}^{N}}\mathrm{d}^{N}\Omega_{N}\:
\big|\Psi_{n_{\theta}L\eta}(\Omega_{N})\big|^{2}=1,
\label{3.16}
\end{equation}
where $\mathrm{d}^{N}\Omega_{N}$ is an infinitesimal surface element
of the unit hypersphere $\mathbb{S}^{N}$. By virtue of the identity
\begin{equation}
\mathrm{d}^{N}\Omega_{N}=\sin^{N-1}\!\theta\,\mathrm{d}\theta\,
\mathrm{d}^{N-1}\Omega_{N-1}
\label{3.17}
\end{equation}
and the normalization condition (\ref{2.9}) for the hyperspherical
harmonics, Eq.\ (\ref{3.16}) yields the integral constraint
\begin{equation}
R^{N}\int_{0}^{\pi}\mathrm{d}\theta\:\sin^{N-1}\!\theta\,
|F_{n_{\theta}L}(\theta)|^{2}=1.
\label{3.18}
\end{equation}
On plugging Eq.\ (\ref{3.15}) into the left-hand side of Eq.\
(\ref{3.18}) and invoking the formula \cite[p.\/~212]{Magn66}
\begin{align}
\int_{-1}^{1}\mathrm{d}x\:(1-x)^{\alpha}(1+x)^{\beta}
P_{n}^{(\alpha,\beta)}(x)P_{n'}^{(\alpha,\beta)}(x)
&= \frac{2^{\alpha+\beta+1}\Gamma(n+\alpha+1)\Gamma(n+\beta+1)}
{n!(2n+\alpha+\beta+1)\Gamma(n+\alpha+\beta+1)}\,\delta_{nn'}
\nonumber \\
& \hspace*{10em} (\textrm{$\alpha>-1$, $\beta>-1$}),
\label{3.19}
\end{align}
we finally obtain that, up to an arbitrary phase factor which we set
equal to the unity, the normalized quasi-radial eigenfunctions,
expressed in terms of the Jacobi polynomials, are
\begin{subequations}
\begin{align}
F_{n_{\theta}L}(\theta)
&= \sqrt{\frac{n_{\theta}!(2n_{\theta}+\mu_{L1}+\mu_{L2}+1)
\Gamma(n_{\theta}+\mu_{L1}+\mu_{L2}+1)}
{R^{N}2^{\mu_{L1}+\mu_{L2}+1}\Gamma(n_{\theta}+\mu_{L1}+1)
\Gamma(n_{\theta}+\mu_{L2}+1)}}
\nonumber \\
& \quad \times(1-\cos\theta)^{\mu_{L2}/2-N/4+1/2}
(1+\cos\theta)^{\mu_{L1}/2-N/4+1/2}
P_{n_{\theta}}^{(\mu_{L2},\mu_{L1})}(\cos\theta),
\label{3.20a}
\end{align}
or equivalently
\begin{align}
F_{n_{\theta}L}(\theta)
&= \sqrt{\frac{n_{\theta}!(2n_{\theta}+\mu_{L1}+\mu_{L2}+1)
\Gamma(n_{\theta}+\mu_{L1}+\mu_{L2}+1)}
{R^{N}2^{N-1}\Gamma(n_{\theta}+\mu_{L1}+1)
\Gamma(n_{\theta}+\mu_{L2}+1)}}
\nonumber \\
& \quad \times\left(\sin\frac{\theta}{2}\right)^{\mu_{L2}-N/2+1}
\left(\cos\frac{\theta}{2}\right)^{\mu_{L1}-N/2+1}
P_{n_{\theta}}^{(\mu_{L2},\mu_{L1})}(\cos\theta).
\label{3.20b}
\end{align}
\label{3.20}%
\end{subequations}
If Eqs.\ (\ref{3.20b}) and (\ref{3.14}) are combined to yield the
normalized $F_{n_{\theta}L}(\theta)$ in terms of the hypergeometric
function ${}_{2}F_{1}$ rather than in terms of the Jacobi polynomial,
in the case of $N=2$ the resulting expression may be shown to
coincide, up to notational differences and up to the factor $R^{-1}$
due to slightly different normalization conditions used, with what
follows from Eqs.\ (22) and (24) of Ref.\ \cite{Kaza13}.
%
%
\section{Special cases}
\label{IV}
\setcounter{equation}{0}
With the solutions corresponding to the general form of the potential
(\ref{2.11a}) being found, we proceed to the analysis of some
particular cases when the functional form of $V(\theta)$ simplifies.
\subsection{The case of $\omega_{1}=\omega$ and $\omega_{2}=0$}
\label{IV.1}
The potential $V(\theta)$ is then
\begin{equation}
V(\theta)=2m\omega^{2}R^{2}\tan^{2}\!\frac{\theta}{2}
\qquad (0\leqslant\theta\leqslant\pi)
\label{4.1}
\end{equation}
and is the one considered by Mardoyan and Petrosyan \cite{Mard13}.
From Eqs.\ (\ref{3.12}), we obtain
\begin{subequations}
\begin{align}
E_{n_{\theta}L} &= \frac{\hbar^{2}}{2mR^{2}}
\left[n_{\theta}+\frac{L}{2}+\frac{3N}{4}-\frac{1}{2}
+\frac{1}{2}\sqrt{\left(L+\frac{N}{2}-1\right)^{2}
+\left(\frac{4m\omega R^{2}}{\hbar}\right)^{2}}\,\right]
\nonumber \\
& \quad \times\left[n_{\theta}+\frac{L}{2}-\frac{N}{4}+\frac{1}{2}
+\frac{1}{2}\sqrt{\left(L+\frac{N}{2}-1\right)^{2}
+\left(\frac{4m\omega R^{2}}{\hbar}\right)^{2}}\,\right]
-2m\omega^{2}R^{2}
\label{4.2a},
\end{align}
or equivalently
\begin{align}
E_{n_{\theta}L}
&= \frac{\hbar^{2}}{2mR^{2}}
\Bigg[\left(n_{\theta}+\frac{L}{2}+\frac{3N}{4}-\frac{1}{2}\right)
\left(n_{\theta}+\frac{L}{2}-\frac{N}{4}+\frac{1}{2}\right)
+\frac{1}{4}\left(L+\frac{N}{2}-1\right)^{2}
\nonumber \\
& \quad +\left(n_{\theta}+\frac{L}{2}+\frac{N}{4}\right)
\sqrt{\left(L+\frac{N}{2}-1\right)^{2}
+\left(\frac{4m\omega R^{2}}{\hbar}\right)^{2}}\,\Bigg].
\label{4.2b}
\end{align}
\label{4.2}%
\end{subequations}
Save for notational differences, the expression for $E_{n_{\theta}L}$
given in Eq.\ (\ref{4.2b}) is identical to the one displayed in Eq.\
(8) of Ref.\ \cite{Mard13}. Furthermore, in the case discussed here
our Eq.\ (\ref{3.20b}) simplifies to
\begin{align}
F_{n_{\theta}L}(\theta) &= \sqrt{\frac{n_{\theta}!
\left(2n_{\theta}+L+\frac{N}{2}+\mu_{L}\right)
\Gamma\!\left(n_{\theta}+L+\frac{N}{2}+\mu_{L}\right)}
{R^{N}2^{N-1}\Gamma\!\left(n_{\theta}+L+\frac{N}{2}\right)
\Gamma(n_{\theta}+\mu_{L}+1)}}
\nonumber \\
& \quad \times\left(\sin\frac{\theta}{2}\right)^{L}
\left(\cos\frac{\theta}{2}\right)^{\mu_{L}-N/2+1}
P_{n_{\theta}}^{(L+N/2-1,\mu_{L})}(\cos\theta),
\label{4.3}
\end{align}
with
\begin{equation}
\mu_{L}=\sqrt{\left(L+\frac{N}{2}-1\right)^{2}
+\left(\frac{4m\omega R^{2}}{\hbar}\right)^{2}}.
\label{4.4}
\end{equation}
After being combined with Eq.\ (\ref{3.14}), the quasi-radial
eigenfunction in Eq.\ (\ref{4.3}) coincides, again up to notational
differences, with the corresponding expression resulting from Eqs.\
(7) and (9) in Ref.\ \cite{Mard13}.
\subsection{The case of $\omega_{1}=0$ and $\omega_{2}=\omega$}
\label{IV.2}
This case corresponds to the potential $V(\theta)$ of the form
\begin{equation}
V(\theta)=2m\omega^{2}R^{2}\cot^{2}\!\frac{\theta}{2}
\qquad (0\leqslant\theta\leqslant\pi).
\label{4.5}
\end{equation}
The expressions for $E_{n_{\theta}L}$ resulting from Eqs.\
(\ref{3.12}) are the same as in Eqs.\ (\ref{4.2}), whereas Eq.\
(\ref{3.20b}) yields
\begin{align}
F_{n_{\theta}L}(\theta) &= \sqrt{\frac{n_{\theta}!
\left(2n_{\theta}+L+\frac{N}{2}+\mu_{L}\right)
\Gamma\!\left(n_{\theta}+L+\frac{N}{2}+\mu_{L}\right)}
{R^{N}2^{N-1}\Gamma\!\left(n_{\theta}+L+\frac{N}{2}\right)
\Gamma(n_{\theta}+\mu_{L}+1)}}
\nonumber \\
& \quad \times\left(\sin\frac{\theta}{2}\right)^{\mu_{L}-N/2+1}
\left(\cos\frac{\theta}{2}\right)^{L}
P_{n_{\theta}}^{(\mu_{L},L+N/2-1)}(\cos\theta).
\label{4.6}
\end{align}
Since Eq.\ (\ref{4.5}) results from Eq.\ (\ref{4.1}) after the
replacement $\theta\to\pi-\theta$ is made in the latter, the function
(\ref{4.6}) coincides, up to an unimportant phase factor, with the one
that emerges once the same replacement is made on the right-hand side
of Eq.\ (\ref{4.3}) and then the use is made of the identity
\cite[p.\/~210]{Magn66}
\begin{equation}
P_{n}^{(\alpha,\beta)}(-x)=(-1)^{n}P_{n}^{(\beta,\alpha)}(x).
\label{4.7}
\end{equation}
\subsection{The case of $\omega_{1}=\omega_{2}=\omega$}
\label{IV.3}
In this case the potential (\ref{2.11}) reduces to the form
\begin{equation}
V(\theta)=\frac{8m\omega^{2}R^{2}}{\sin^{2}\!\theta}
-4m\omega^{2}R^{2}
\qquad (0\leqslant\theta\leqslant\pi).
\label{4.8}
\end{equation}
Then the energy eigenvalues (\ref{3.12}) become
\begin{subequations}
\begin{align}
E_{n_{\theta}L}
&= \frac{\hbar^{2}}{2mR^{2}}
\left[n_{\theta}+\frac{N}{2}
+\sqrt{\left(L+\frac{N}{2}-1\right)^{2}
+\left(\frac{4m\omega R^{2}}{\hbar}\right)^{2}}\,\right]
\nonumber \\
& \quad \times\left[n_{\theta}-\frac{N}{2}+1
+\sqrt{\left(L+\frac{N}{2}-1\right)^{2}
+\left(\frac{4m\omega R^{2}}{\hbar}\right)^{2}}\,\right]
-4m\omega^{2}R^{2},
\label{4.9a}
\end{align}
or equivalently
\begin{align}
E_{n_{\theta}L}
&= \frac{\hbar^{2}}{2mR^{2}}
\Bigg[\left(n_{\theta}+\frac{N}{2}\right)
\left(n_{\theta}-\frac{N}{2}+1\right)
+\left(L+\frac{N}{2}-1\right)^{2}
\nonumber \\
& \quad +(2n_{\theta}+1)\sqrt{\left(L+\frac{N}{2}-1\right)^{2}
+\left(\frac{4m\omega
R^{2}}{\hbar}\right)^{2}}\,\Bigg]+4m\omega^{2}R^{2}.
\label{4.9b}
\end{align}
\label{4.9}%
\end{subequations}
The quasi-radial eigenfunctions (\ref{3.20}) are
\begin{equation}
F_{n_{\theta}L}(\theta)
=\sqrt{\frac{n_{\theta}!(2n_{\theta}+2\mu_{L}+1)
\Gamma(n_{\theta}+2\mu_{L}+1)}
{R^{N}2^{2\mu_{L}+1}\Gamma^{2}(n_{\theta}+\mu_{L}+1)}}
\left(\sin\theta\right)^{\mu_{L}-N/2+1}
P_{n_{\theta}}^{(\mu_{L},\mu_{L})}(\cos\theta),
\label{4.10}
\end{equation}
with $\mu_{L}$ defined as in Eq.\ (\ref{4.4}). By virtue of the
relationship \cite[p.\/~219]{Magn66}
\begin{equation}
P_{n}^{(\mu,\mu)}(x)
=\frac{2^{2\mu}\Gamma\big(\mu+\frac{1}{2}\big)\Gamma(n+\mu+1)}
{\sqrt{\pi}\,\Gamma(n+2\mu+1)}\,C_{n}^{\mu+1/2}(x),
\label{4.11}
\end{equation}
linking the Jacobi polynomial $P_{n}^{(\mu,\mu)}(x)$ to the Gegenbauer
polynomial $C_{n}^{\mu+1/2}(x)$, $F_{n_{\theta}L}(\theta)$ may be cast
into the form
\begin{equation}
F_{n_{\theta}L}(\theta)
=\sqrt{\frac{2^{2\mu_{L}-1}n_{\theta}!(2n_{\theta}+2\mu_{L}+1)
\Gamma^{2}\big(\mu_{L}+\frac{1}{2}\big)}
{R^{N}\pi\,\Gamma(n_{\theta}+2\mu_{L}+1)}}
\left(\sin\theta\right)^{\mu_{L}-N/2+1}
C_{n_{\theta}}^{\mu_{L}+1/2}(\cos\theta).
\label{4.12}
\end{equation}

In the limit $\omega\to0$, corresponding to the case of a free
particle on $\mathbb{S}_{R}^{N}$, Eqs.\ (\ref{4.9}) and (\ref{4.12})
go over, as they should, into
\begin{equation}
E_{n_{\theta}L}
=\frac{\hbar^{2}}{2mR^{2}}(n_{\theta}+L)(n_{\theta}+L+N-1)
\label{4.13}
\end{equation}
and
\begin{equation}
F_{n_{\theta}L}(\theta)
=\sqrt{\frac{2^{2L+N-3}n_{\theta}!(2n_{\theta}+2L+N-1)
\Gamma^{2}\big(L+\frac{N-1}{2}\big)}
{R^{N}\pi\,\Gamma(n_{\theta}+2L+N-1)}}\,
\sin^{L}\!\theta\,C_{n_{\theta}}^{L+N/2-1/2}(\cos\theta),
\label{4.14}
\end{equation}
respectively.
%
%
\section{The Euclidean limit $(R\to\infty)$}
\label{V}
\setcounter{equation}{0}
\subsection{General considerations}
\label{V.1}
Consider a Schr{\"o}dinger particle moving on the hypersphere
$\mathbb{S}_{R}^{N}$ in a reasonably arbitrary longitudinal potential
$V(\theta)$. On the route to the Euclidean limit for this system, we
first project stereographically the hypersphere $\mathbb{S}_{R}^{N}$
from its \emph{south\/} pole onto the hyperplane tangent to
$\mathbb{S}_{R}^{N}$ at the \emph{north\/} pole [with regard to the
nomenclature used here, cf.\ the pertinent remark under Eq.\
(\ref{2.2c})]. In this tangent hyperplane, we introduce the
hyperspherical coordinates $r,\theta_{N-1},\ldots,\theta_{1}$. The
angles $\{\theta_{n}\}_{n=1}^{N-1}$ are defined as in Sec.\ \ref{II},
whereas the radial variable $r$ is related to the hyperlatitude
$\theta\equiv\theta_{N}$ on $\mathbb{S}_{R}^{N}$ through
\begin{equation}
r=2R\tan\frac{\theta}{2}
\quad
\Rightarrow
\quad
\theta=2\arctan\frac{r}{2R}.
\label{5.1}
\end{equation}
Clearly, it holds that
\begin{equation}
\theta\stackrel{R\to\infty}{\rightarrow}\frac{r}{R}.
\label{5.2}
\end{equation}
Next we define
\begin{equation}
U(r)=V(\theta)
\label{5.3}
\end{equation}
and put\footnote{~For the general potential $V(\theta)$ considered
here, the quasi-radial quantum number $n_{\theta}$ equals the number
of zeros of the eigenfunction $F_{n_{\theta}L}(\theta)$ in the
\emph{open\/} interval
$0<\theta<\pi$.}\textsuperscript{,}\footnote{~The function
$f_{n_{r}L}(r)$ used in this section should not be confused with the
function $f_{L}(\rho)$ that appeared in Eqs.\ (\ref{3.3}), (\ref{3.4})
and (\ref{3.7}).}
\begin{equation}
F_{n_{\theta}L}(\theta)
=\left(1+\frac{r^{2}}{4R^{2}}\right)^{N/2-1}f_{n_{r}L}(r),
\label{5.4}
\end{equation}
where for convenience we define $n_{r}\equiv n_{\theta}$. On inserting
Eqs.\ (\ref{5.3}) and (\ref{5.4}) into the quasi-radial
Schr{\"o}dinger equation (\ref{2.10}) (with $E$ identified with the
energy eigenvalue $E_{n_{r}L}$), we find that the radial function
$f_{n_{r}L}(r)$ solves the equation
\begin{align}
& \Bigg\{\hspace*{-0.25em}-\!\frac{\hbar^{2}}{2m}
\left[\frac{\mathrm{d}^{2}}{\mathrm{d}r^{2}}
+\frac{N-1}{r}\frac{\mathrm{d}}{\mathrm{d}r}
-\frac{L(L+N-2)}{r^{2}}\right]
\nonumber \\
& \qquad +\left(\frac{4R^{2}}{r^{2}+4R^{2}}\right)^{2}
\left[U(r)-\frac{\hbar^{2}N(N-2)}{8mR^{2}}-E_{n_{r}L}\right]
\hspace*{-0.25em}\Bigg\}f_{n_{r}L}(r)=0,
\label{5.5}
\end{align}
the normal (i.e., without the first derivative) form of which is
\begin{align}
& \Bigg\{\hspace*{-0.25em}
-\!\frac{\hbar^{2}}{2m}
\frac{\mathrm{d}^{2}}{\mathrm{d}r^{2}}
+\frac{\hbar^{2}\left(L+\frac{N-3}{2}\right)\!
\left(L+\frac{N-1}{2}\right)}{2mr^{2}}
\nonumber \\
& \qquad +\left(\frac{4R^{2}}{r^{2}+4R^{2}}\right)^{2}
\left[U(r)-\frac{\hbar^{2}N(N-2)}{8mR^{2}}-E_{n_{r}L}\right]
\hspace*{-0.25em}\Bigg\}r^{(N-1)/2}f_{n_{r}L}(r)=0.
\label{5.6}
\end{align}
It is easy to see from Eqs.\ (\ref{5.4}) and (\ref{5.1}) that if
$F_{n_{\theta}L}(\theta)$ is forced to obey the normalization
constraint
\begin{equation}
R^{N}\int_{0}^{\pi}\mathrm{d}\theta\:\sin^{N-1}\!\theta\,
\big|F_{n_{\theta}L}(\theta)\big|^{2}=1
\label{5.7}
\end{equation}
[a special case of which, for the potential (\ref{2.11}), has been
given in Eq.\ (\ref{3.18})], then $f_{n_{r}L}(r)$ is normalized in the
sense of
\begin{equation}
\int_{0}^{\infty}\mathrm{d}r\:r^{N-1}
\left(\frac{4R^{2}}{r^{2}+4R^{2}}\right)^{2}
\big|f_{n_{r}L}(r)\big|^{2}=1.
\label{5.8}
\end{equation}

The presence of the weight function
$\left(\frac{4R^{2}}{r^{2}+4R^{2}}\right)^{2}$ in Eqs.\ (\ref{5.5}),
(\ref{5.6}) and (\ref{5.8}) is a manifestation of the fact that we
still remain outside the realm of the Euclidean geometry. The
transition to the latter is achieved only in the next step, by going
to the infinity with the value of the radius $R$ of the hypersphere.
On defining
\begin{equation}
\widetilde{U}(r)=\lim_{R\to\infty}U(r),
\qquad
\widetilde{E}_{n_{r}L}=\lim_{R\to\infty}E_{n_{r}L},
\qquad
\widetilde{f}_{n_{r}L}(r)=\lim_{R\to\infty}f_{n_{r}L}(r)
\label{5.9}
\end{equation}
and making the limiting passage $R\to\infty$ in Eqs.\ (\ref{5.5}),
(\ref{5.6}) and (\ref{5.8}), we find that they take the desired
Euclidean forms
\begin{equation}
\left\{-\frac{\hbar^{2}}{2m}
\left[\frac{\mathrm{d}^{2}}{\mathrm{d}r^{2}}
+\frac{N-1}{r}\frac{\mathrm{d}}{\mathrm{d}r}
-\frac{L(L+N-2)}{r^{2}}\right]
+\widetilde{U}(r)-\widetilde{E}_{n_{r}L}\right\}
\widetilde{f}_{n_{r}L}(r)=0,
\label{5.10}
\end{equation}
\begin{equation}
\left[-\frac{\hbar^{2}}{2m}\frac{\mathrm{d}^{2}}{\mathrm{d}r^{2}}
+\frac{\hbar^{2}\left(L+\frac{N-3}{2}\right)\!
\left(L+\frac{N-1}{2}\right)}
{2mr^{2}}+\widetilde{U}(r)-\widetilde{E}_{n_{r}L}\right]
r^{(N-1)/2}\widetilde{f}_{n_{r}L}(r)=0
\label{5.11}
\end{equation}
and
\begin{equation}
\int_{0}^{\infty}\mathrm{d}r\:r^{N-1}
\big|\widetilde{f}_{n_{r}L}(r)|^{2}=1,
\label{5.12}
\end{equation}
respectively.
\subsection{Application to the hyperspherical P{\"o}schl--Teller
oscillator}
\label{V.2}
In the case of the hyperspherical P{\"o}schl--Teller oscillator, the
procedure described above will certainly be well-defined if we assume
that of the two strength parameters characterizing the potential, the
parameter $\omega_{1}$ is independent of the radius $R$, while the
parameter $\omega_{2}$ is inversely proportional to $R$ squared. By
making the adequate substitution
\begin{equation}
\omega_{2}=\frac{\hbar\chi}{4mR^{2}},
\label{5.13}
\end{equation}
where $\chi>0$ is independent of $R$, and dropping from now on, for
clarity of notation, the index 1 at $\omega_{1}$, we get
\begin{equation}
V(\theta)=2m\omega^{2}R^{2}\tan^{2}\!\frac{\theta}{2}
+\frac{\hbar^{2}\chi^{2}}{8mR^{2}}\cot^{2}\!\frac{\theta}{2}.
\label{5.14}
\end{equation}
Hence, with the use of Eqs.\ (\ref{5.3}) and (\ref{5.1}), as well as
of the first of Eqs.\ (\ref{5.9}), it follows that
\begin{equation}
U(r)=\frac{m\omega^{2}r^{2}}{2}+\frac{\hbar^{2}\chi^{2}}{2mr^{2}}
=\widetilde{U}(r).
\label{5.15}
\end{equation}
This potential describes the isotropic harmonic oscillator subjected
to an additional centrifugal-like force. With
$\widetilde{U}(r)$ in the form (\ref{5.15}), Eq.\ (\ref{5.11}) becomes
\begin{equation}
\left[-\frac{\hbar^{2}}{2m}\frac{\mathrm{d}^{2}}{\mathrm{d}r^{2}}
+\frac{\hbar^{2}\Lambda_{L}(\Lambda_{L}+1)}{2mr^{2}}
+\frac{m\omega^{2}r^{2}}{2}-\widetilde{E}_{n_{r}L}\right]
r^{(N-1)/2}\widetilde{f}_{n_{r}L}(r)=0,
\label{5.16}
\end{equation}
where
\begin{equation}
\Lambda_{L}=\sqrt{\left(L+\frac{N}{2}-1\right)^{2}+\chi^{2}}
-\frac{1}{2}
\qquad \left(=\mu_{L2}-\frac{1}{2}\right).
\label{5.17}
\end{equation}
Solving Eq.\ (\ref{5.16}) subject to the boundary conditions
\begin{equation}
\textrm{$\widetilde{f}_{n_{r}L}(r)$ bounded for $r\to0$},
\qquad
\widetilde{f}_{n_{r}L}(r)\stackrel{r\to\infty}\longrightarrow0
\label{5.18}
\end{equation}
belongs to the class of standard exercises in intermediate quantum
mechanics (cf., for instance, Ref.\ \cite[Problem~65]{Flug99}). One
finds that
\begin{equation}
\widetilde{E}_{n_{r}L}=\hbar\omega\left[2n_{r}+1
+\sqrt{\left(L+\frac{N}{2}-1\right)^{2}+\chi^{2}}\,\right]
\qquad (n_{r}\in\mathbb{N}_{0})
\label{5.19}
\end{equation}
and, once the normalization constraint (\ref{5.12}) is imposed and
phase factors in $\widetilde{f}_{n_{r}L}(r)$ are suitably adjusted,
that\footnote{~In Ref.\ \cite[Problem~65]{Flug99}, the radial wave
functions were expressed in terms of the confluent geometric function.
In the present work, we prefer to use the generalized Laguerre
polynomials instead.}
\begin{equation}
\widetilde{f}_{n_{r}L}(r)=\sqrt{\frac{2n_{r}!}
{\Gamma\!\left(n_{r}+\Lambda_{L}+\frac{3}{2}\right)}}
\left(\frac{m\omega}{\hbar}\right)^{N/4}
\left(\frac{m\omega r^{2}}{\hbar}\right)^{\Lambda_{L}/2-N/4+3/4}
\mathrm{e}^{-m\omega r^{2}/2\hbar}
L_{n_{r}}^{(\Lambda_{L}+1/2)}\left(\frac{m\omega r^{2}}{\hbar}\right),
\label{5.20}
\end{equation}
where $L_{n}^{(\alpha)}(x)$ is the generalized Laguerre polynomial
\cite[Sec.\/~5.5]{Magn66}.

Of course, one should expect to be able to arrive at Eq.\ (\ref{5.19})
also in another way, namely after combining Eqs.\ (\ref{3.12}) and
(\ref{5.13}), and taking then the limit $R\to\infty$; it presents no
difficulties to show that this is indeed the case. Similarly, Eq.\
(\ref{5.20}) should follow (possibly up to a sign factor) from Eqs.\
(\ref{5.4}) and (\ref{3.20}), after the aforementioned limit is taken.
However, in this case the proof appears to be a bit more complex. On
exploiting Eqs.\ (\ref{5.4}), (\ref{3.20b}), (\ref{5.1}) and
(\ref{5.17}), we may write
\begin{align}
f_{n_{r}L}(r) &= \sqrt{\frac{n_{r}!
\left(2n_{r}+\mu_{L1}+\Lambda_{L}+\frac{3}{2}\right)
\Gamma\!\left(n_{r}+\mu_{L1}+\Lambda_{L}+\frac{3}{2}\right)}
{R^{N}2^{N-1}\Gamma(n_{r}+\mu_{L1}+1)
\Gamma\!\left(n_{r}+\Lambda_{L}+\frac{3}{2}\right)}}
\nonumber \\
& \quad \times
\frac{\displaystyle\left(\frac{r}{2R}\right)^{\Lambda_{L}-N/2+3/2}}
{\displaystyle
\left(1+\frac{r^{2}}{4R^{2}}\right)^{\mu_{L1}/2+\Lambda_{L}/2+1/4}}
P_{n_{r}}^{(\Lambda_{L}+1/2,\mu_{L1})}
\left(\frac{\displaystyle1-\frac{r^{2}}{4R^{2}}}
{\displaystyle1+\frac{r^{2}}{4R^{2}}}\right).
\label{5.21}
\end{align}
Now, as it holds that
\begin{equation}
\mu_{L1}\stackrel{R\to\infty}{\longrightarrow}
\frac{4m\omega R^{2}}{\hbar}
\label{5.22}
\end{equation}
[cf.\ Eq.\ (\ref{3.6})], we obviously have
\begin{equation}
\lim_{R\to\infty}
\left(1+\frac{r^{2}}{4R^{2}}\right)^{\mu_{L1}/2+\Lambda_{L}/2+1/4}
=\mathrm{e}^{m\omega r^{2}/2\hbar}.
\label{5.23}
\end{equation}
Furthermore, from Eq.\ (\ref{5.22}) and from the following two
asymptotic relationships:
\begin{equation}
\frac{\Gamma(x+a)}{\Gamma(x+b)}
\stackrel{x\to\infty}{\longrightarrow}x^{a-b}
\label{5.24}
\end{equation}
(cf.\ Ref.\ \cite[p.\/~12]{Magn66}) and
\begin{equation}
\lim_{\beta\to\infty}
P_{n}^{(\alpha,\beta)}\left(1-\frac{2x}{\beta}\right)
=L_{n}^{(\alpha)}(x)
\label{5.25}
\end{equation}
(cf.\ Ref.\ \cite[p.\/~247]{Magn66}), it follows that
\begin{equation}
\frac{\Gamma\!\left(n_{r}+\mu_{L1}+\Lambda_{L}+\frac{3}{2}\right)}
{\Gamma(n_{r}+\mu_{L1}+1)}\stackrel{R\to\infty}{\longrightarrow}
\left(\frac{4m\omega R^{2}}{\hbar}\right)^{\Lambda_{L}+1/2}
\label{5.26}
\end{equation}
and
\begin{equation}
\lim_{R\to\infty}P_{n_{r}}^{(\Lambda_{L}+1/2,\mu_{L1})}
\left(\frac{\displaystyle1-\frac{r^{2}}{4R^{2}}}
{\displaystyle1+\frac{r^{2}}{4R^{2}}}\right)
=\lim_{R\to\infty}P_{n_{r}}^{(\Lambda_{L}+1/2,\mu_{L1})}
\left(1-\frac{r^{2}}{2R^{2}}\right)
=L_{n_{r}}^{(\Lambda_{L}+1/2)}
\left(\frac{m\omega r^{2}}{\hbar}\right),
\label{5.27}
\end{equation}
respectively. On taking the limit $R\to\infty$ on the right-hand side
of Eq.\ (\ref{5.21}) and employing then Eqs.\ (\ref{5.23}),
(\ref{5.26}) and (\ref{5.27}), after some algebra one indeed arrives
at Eq.\ (\ref{5.20}).
\section{Conclusions}
\label{VI}
In this work, we have arrived at analytical solutions to an energy
eigenvalue problem for a Schr{\"o}dinger particle moving on an
($N\geqslant2$)-dimensional hypersphere $\mathbb{S}_{R}^{N}$ in the
field of the longitudinal potential
$V(\theta)=2m\omega_{1}^{2}R^{2}\tan^{2}(\theta/2)
+2m\omega_{2}^{2}R^{2}\cot^{2}(\theta/2)$. This system may be viewed
as an $N$-dimensional generalization of the one-dimensional anharmonic
trigonometric oscillator of P{\"o}schl and Teller \cite{Posc33}.

In the special case of $N=2$, our results coincide with those
presented in Ref.\ \cite{Kaza13} by Kazaryan \emph{et al}. On the
other hand, in the limit $\omega_{2}=0$ our findings reproduce those
derived by Mardoyan and Petrosyan \cite{Mard13} for their model of an
isotropic harmonic oscillator on $\mathbb{S}_{R}^{N}$.

If the parameter $\omega_{2}$ in $V(\theta)$ depends on the radius of
$\mathbb{S}_{R}^{N}$ in the inverse-square manner and if $\omega_{1}$
is independent of $R$, then the Euclidean limit for the system
considered here does exist. This limit may be achieved by performing a
suitable stereographic projection of $\mathbb{S}_{R}^{N}$ onto a
tangent space, followed by the passage with $R$ to infinity. In
result, our hyperspherical P{\"o}schl--Teller oscillator goes over
into an isotropic harmonic oscillator in $\mathbb{R}^{N}$ perturbed by
a centrifugal-type force.
%
%
\section*{Acknowledgments}
Access to Mathematica software, granted by CI TASK, is acknowledged.
\section*{Author Declarations}
The author has no conflicts to disclose.
%
%

%
\end{document}